\documentclass[11pt]{article}\usepackage[dvipsnames,usenames]{color}

\begin{document}
\begin{normalsize}
\title{Hodograph transformations for a Camassa-
Holm hierarchy\\ in $2+1$ dimensions}
\author{P. G. Est\'evez $^{(a)}$ and J. Prada $^{(b)}$\\ $^{(a)}$
Area de Fisica Te\'orica.\\
Universidad de Salamanca, Salamanca 37008, Spain\\
\textmd{{ pilar@usal.es}}\\ $^{(b)}$ Departamento de Matem\'aticas.\\
Universidad de Salamanca, Salamanca 37008, Spain\\
\textmd{{ prada@usal.es}}} \end{normalsize}

\date{}
\maketitle

\begin{abstract}
A generalization of the negative Camassa-Holm hierarchy to $2+1$ dimensions is presented under the name
CHH(2+1). Several hodograph transformations are applied in order to transform the hierarchy into a system of
coupled CBS (Calogero-Bogoyavlenskii-Schiff) equations in $2+1$ dimensions that pass the Painlev\'e test. A
non-isospectral Lax pair for CHH(2+1) is obtained through the above mentioned relationship with the CBS
spectral problem.

\end{abstract}

\section{Introduction}\label{intro}
\setcounter{equation}{0}

The seminal papers in which the Camassa-Holm equation was described \cite{ch93}, \cite{chh94} have led to a
much work related to equations with peakon solutions. In particular in references \cite{gp95}, \cite{dhh02}
and \cite{hw03}, the authors include the Camassa-Holm equation within a wider class of equations with
peakons. The integrability of the Camassa-Holm equation, spectral problem, solutions, etc have been studied
in many papers in the last ten years (See, for instance, references \cite{ch93}, \cite{dhh02} and
\cite{hw03}).

 The Painlev\'e test \cite{Weiss} is usually presented as a powerful instrument to check the
integrability of an equation. Nevertheless, in \cite{gp95} the limitations of the Painlev\'e test when
applied to Camassa-Holm like equations are discussed.

The Painlev\'e property provides not only the basis for the Painlev\'e test, but also for the Singular
Manifold Method \cite{Weiss}. When an equation passes the Painlev\'e test, the singular manifold method can
be applied to  algorithmically construct the Lax pair \cite{eh00}, \cite{e04} and many other properties of
the integrable systems such as Darboux transformations, $\tau$-functions, etc. The main problem with
Painlev\'e methods is that the Painlev\'e property is non invariant under changes of independent and/or
dependent variables. Often, finding the change of variables that writes an equation in a form that passes the
Painlev\'e test,  is a question of luck or ability.

From the point of view of the spectral problem, the Lax pair for a partial differential equation is usually
found by inspection. Most frequently, a spectral problem is proposed and then the equations that satisfy this
spectral problem are derived \cite{ma80}, \cite{AF88}, \cite{cgp97}. In contrast, the Singular Manifold
Method has the attractive property that it allows us to start from a given equation (that passes the
Painlev\'e test) and derive its Lax pair in a very precise way. Our conjecture is that if an equation is
integrable, there must be a transformation that will let us  transform the equation into a new one in which
the Painlev\'e test is successful and the Singular Manifold Method can be applied to derive the Lax pair.

In \cite{dhh02} \cite{h00}, hodograph transformations were proposed  as useful instruments to transform
peakon equations into equations that pass the Painlev\'e test. Based on this idea,  in section 2 of this
paper we attempt to study the integrability of a $n$-component Camassa-Holm hierarchy in $2+1$ dimensions
(which we will call CHH(2+1)) by means of several hodograph transformations that map this hierarchy in a
system of $n$ coupled CBS (Calogero-Bogoyavlenskii-Schiff \cite{bog90}, \cite{cal}) equations in $3$
independent variables that are different for each CBS component. This result generalizes those obtained in
\cite{h00} where reciprocal transformations between  the first component of the CHH(2+1) and CBS are studied.

The CBS equation in $3$ dimensions has been proved to pass the Painlev\'e test \cite{e04}. In the same
reference, the Singular Manifold Method was used to construct the Lax pair, which in fact is a
non-isospectral one \cite{cgp97}, \cite{e04}. This knowledge of the spectral problem associated with the CBS
equation allows us to devote section 3 to reversing the hodograph transformations and rewrite the spectral
problem in the original variables. Thus, a non-isospectral Lax pair for the CHH(2+1) hierarchy is obtained.
The coincidences and differences  between these results and other spectral problems are discussed at the end
of this section.

The conclusions are presented in section 4.

\section{Camassa-Holm Hierarchy in 2+1 dimensions}\label{2}
\setcounter{equation}{0} \noindent $\bullet$ As  is well known \cite{hq02}, the negative Camassa-Holm
hierarchy for a field $u(x,t)$ can be written as:
\begin{equation}
u_t=R^{-n}u_x,\qquad R=J_0J_1^{-1}\qquad n\geq 1,\label{2.1}
\end{equation}
where $n$ is an integer number that is the order of the hierarchy and $J_0$, $J_1$ are the following
operators:
\begin{equation}
 \qquad J_0=(\partial^3-\partial), \qquad J_1=(u\partial+\partial u),\qquad \partial=\frac{\partial}{\partial x}\quad.\label{2.2}
\end{equation}
For our purpose it is convenient to introduce $n$ functions $v_1(x,t)...v_n(x,t)$ defined as:

\begin{eqnarray}
v_1&=&J_0^{-1}u_x\Longrightarrow J_0v_1=u_x\nonumber\\ v_j&=&J_0^{-1}J_1v_{j-1}\Longrightarrow
J_0v_j=J_1v_{j-1},\qquad \qquad j=2...n,\quad.\label{2.3}
\end{eqnarray}
Equation (\ref{2.1}) can be written now simply as:
\begin{equation}
u_t=J_1v_{n}\quad,\label{2.4}
\end{equation}
and hence the negative Camassa-Holm hierarchy can be considered as the $n+1$ equations
(\ref{2.3})-(\ref{2.4}) in $n+1$ fields $u, v_1,...v_n$.

 Obviously, for $n=1$, the system (\ref{2.3})-(\ref{2.4}) reduces to:
\begin{eqnarray} u_t&=&2u(v_1)_x+u_xv_1\nonumber\\u&=&(v_{1})_{xx}-v_{1} \quad,\label{2.5}\end{eqnarray}
which is the celebrated Camassa-Holm equation \cite{chh94}.

\medskip

\noindent $\bullet$ The positive Camassa-Holm hierarchy \cite{hq02} would be obtained through
\begin{equation}
u_t=R^{n}(0),\quad n\geq 1,\label{2.6}
\end{equation}
whose $n=1$ component is:
$$u_t=J_0J_1^{-1}(0)\quad;$$
or equivalently: $$ u_t=J_0 v_1$$ $$ J_1 v_1=0\Longrightarrow v_1=u^{-1/2}\quad,$$ which is the Dym equation
\cite{k75} with an extra term $(v_1)_x$ \cite{cgp97}, \cite{AF88}.

\subsection{Generalization to $3$ dimensions}
A simple generalization of (\ref{2.3})-(\ref{2.4}) to $3$ dimensions is as follows:
\begin{eqnarray}
U_y&=&J_0V_1\nonumber\\  J_0V_j&=&J_1V_{j-1},\qquad \qquad j=2...n\quad,\label{2.7}\\U_t&=&J_1V_{n}\nonumber
\end{eqnarray}
 where $U=U(x,t,y)$, $V_j=V_j(x,t,y)$.

 System (\ref{2.7}) can also be written as:
\begin{equation}
U_t=R^{-n}U_y \quad.\label{2.8}
\end{equation}
The equivalent positive hierarchy should be:

\begin{equation}
U_t=R^{n}U_y\quad,\label{2.9}
\end{equation}
which can be trivially obtained from (2.8) by interchanging $t$ and $y$. Consequently, in $3$ dimensions
(2.8) contains both the negative and positive hierarchies. One can be obtained from the other by
interchanging the roles of $t$ and $y$.

It is also necessary to point out that the first component of (\ref{2.7}) can be written (by simply doing
$V_1=m_y$) as:
$$(\partial_t-2m_{xy}-m_y\partial_x)(m_{xxx}-m_x)=0$$
that is a generalization to $2+1$ dimensions of the Fokas-Fuchssteiner-Camassa-Holm  equation \cite{ff81}
proposed in \cite{cgp97} and analyzed in \cite{h00}.

\subsubsection*{Reductions}

 \noindent $\bullet$ It is trivial to see that the negative Camassa-Holm hierarchy (\ref{2.1}) would be obtained from
  (2.8) through the reduction $\frac{\partial}{\partial y}=\frac{\partial}{\partial x}$.

\medskip

  \noindent $\bullet$ If we reduce (\ref{2.8}) by
  setting $\frac{\partial}{\partial t}=0$ we obtain:
  $$R^{-n}U_y=0\Longrightarrow U_y=R^n(0)$$
 which is the positive hierarchy (2.6) where $t$ has been replaced by $y$.

Notice that (\ref{2.8}) is formally included in the Dym case of reference \cite{cgp97}. Nevertheless, the
generalization of the Camassa-Holm hierarchy that the authors construct explicitly in that work is not
(\ref{2.9}) because it corresponds to $n=1$ and $U$ is a field with $N$ components (see equation (2.16) of
this reference). Only  the first component of both hierarchies ($n=N=1$) coincides.

Below we shall denote (\ref{2.7}) by CHH(2+1) and we   prove through several hodograph transformations that
it can be transformed into a system that passes the Painlev\'e test.
\subsection{First Hodograph transformation}
If we set:
\begin{equation}
U=P^2\quad,\label{2.10}
\end{equation}
we can write  system (\ref{2.7}) as:
\begin{eqnarray}
P_y&=&(\beta_1)_x\label{2.11}\\ \frac{J_0V_j}{2P} &=&(PV_{j-1})_x,\qquad \qquad j=2...n\quad ,\label{2.12}\\
P_t&=&(PV_n)_x\label{2.13}
\end{eqnarray}
where we have defined:\begin{equation}(\beta_1)_x= \frac{J_0V_1}{2P}\quad .\label{2.14}
\end{equation}
The conservative form of (\ref{2.11}) and (\ref{2.13}) allows us, according to \cite{h00} and \cite{dhh02},
to define the following  hodograph transformation:
\begin{eqnarray} dX&=&Pdx+PV_ndt+\beta_1dy\nonumber\\ Z_1&=&t \label{2.15}\\ Y&=&y \quad ,\nonumber
\end{eqnarray}
The partial derivatives are now:  \begin{equation}\frac{\partial}{\partial x}=P\frac{\partial}{\partial
X},\quad \frac{\partial}{\partial t}=\frac{\partial}{\partial Z_1}+PV_n \frac{\partial}{\partial X},\quad
\frac{\partial}{\partial y}=\frac{\partial}{\partial Y}+\beta_1 \frac{\partial}{\partial X}\label{2.16}\quad
.\end{equation} The inverses of (\ref{2.15})-(\ref{2.16}) are:
\begin{eqnarray} dx&=&\frac{dX}{P}-V_ndZ_1-\frac{\beta_1}{P}dY\nonumber\\ t&=&Z_1\\ y&=&Y \nonumber \quad ,\label{2.17}
\end{eqnarray}
\begin{equation}\frac{\partial}{\partial X}=\frac{1}{P}\frac{\partial}{\partial x},\quad
\frac{\partial}{\partial Z_1}=\frac{\partial}{\partial t}-V_n \frac{\partial}{\partial x},\quad
\frac{\partial}{\partial Y}=\frac{\partial}{\partial y}-\frac{\beta_1}{P} \frac{\partial}{\partial x} \quad .
\label{2.18}\end{equation} With this hodograph transformation, system (\ref{2.11})-(\ref{2.14}) becomes:
\begin{eqnarray}
P_Y&=&P(\beta_1)_X-P_X\beta_1\label{2.19}\\\frac{P_{Z_1}}{P^2}&=&(V_n)_X\label{2.20}\\
\frac{1}{2P}\left(\left\{P\left[P(V_j)_X\right]_X\right\}_X-(V_j)_X\right)&=&(PV_{j-1})_X, \qquad j=2...n\label{2.21}\\
\frac{1}{2P}\left(\left\{P\left[P(V_1)_X\right]_X\right\}_X-(V_1)_X\right)&=&(\beta_1)_X\qquad \label{2.22}
\end{eqnarray}
Nevertheless,  (\ref{2.19})-(\ref{2.22}) is not yet a system in which the Lax pair can be directly derived. A
new set of transformations are needed in order to write  (\ref{2.19})-(\ref{2.22}) in a form in which the
singular manifold method could be applied to derive the Lax pair.

\subsection{Second Hodograph transformation}

\noindent $\bullet$ Let us take (\ref{2.21}) for $j=n$:
$$\frac{1}{2P}\left(\left\{P\left[P(V_n)_X\right]_X\right\}_X-(V_n)_X\right)=(PV_{n-1})_X\quad ,$$
and by substituting  (\ref{2.20}), the result is:
\begin{equation}
\left(\frac{P_{XX}}{2P}+\frac{1-P_X^2}{4P^2}\right)_{Z_1}=\left(PV_{n-1}\right)_X\quad .\label{2.23}
\end{equation}
The form of equation  (\ref{2.23}) suggests that we should introduce  a new function $H$, defined as
\begin{equation}
H_X=\left(\frac{P_{XX}}{2P}+\frac{1-P_X^2}{4P^2}\right)\quad ,\label{2.24}
\end{equation}
which allows us to integrate  (\ref{2.23}) as:
\begin{equation}
PV_{n-1}=H_{Z_1}\quad .\label{2.25}
\end{equation}

\medskip

\noindent $\bullet$ Based on (\ref{2.25}), let us introduce  $Z_2...Z_{n-1}$ new independent variables, such
that equation (\ref{2.25}) can be extended through the following definition:
\begin{equation}
PV_{n-j}=H_{Z_j},\quad j=2...n-1\Longrightarrow PV_j=H_{Z_{n-j}},\quad j=1...n-2\quad .\label{2.26}
\end{equation}
Notice that (\ref{2.26}) are hodograph transformations between each dependent variable $V_j$ and the
corresponding independent variable $Z_{n-j}$.

\medskip

\noindent $\bullet$ Taking (\ref{2.21}) para $n-j$:
$$\frac{1}{2P}\left(\left\{P\left[P(V_{n-j})_X\right]_X\right\}_X-(V_{n-j})_X\right)=(PV_{n-j-1})_X,\quad j=1...n-2\quad ,$$
and by using (\ref{2.26})
\begin{equation}\frac{1}{2P}\left(\left\{P\left[P\left(\frac{H_{Z_j}}{P}\right)_X\right]_X\right\}_X-\left(\frac{H_{Z_j}}
{P}\right)_X\right)=\left(H_{Z_{j+1}}\right)_X\quad .\label{2.27}\end{equation} We can use (\ref{2.24}) to
obtain:
\begin{equation}
P_{XX}=2PH_X+\frac{P_X^2-1}{2P}\quad .\label{2.28}
\end{equation}
By substituting (\ref{2.28}) in (\ref{2.27}), we have:
\begin{equation}
H_{XXXZ_j}-4H_{XZ_j}H_{X}-2H_{XX}H_{Z_j}=2H_{XZ_{j+1}},\quad j=1...n-2\quad .\label{2.29}
\end{equation}
Each of the equations of (\ref{2.29}) is a CBS equation in the $3$ variables $X$, $Z_j$ and $Z_{j+1}$. This
equation   has been studied by different authors (see \cite{cal}, \cite {bog90}, \cite{cgp97}, \cite{e04},
\cite{h00}). This equation can be also considered as a generalization to $2+1$ dimensions of the AKNS
(Ablowitz, Kaup, Neweel, Segur) equation. Its Lax pair can be found though the singular manifold method in
\cite{e04} and it proves to be non-isospectral \cite{cgp97}, \cite{kp98}, \cite{e04}. We shall use this
result in the next section.

\subsection{Third Hodograph transformation}

\noindent $\bullet$ By substituting $V_1=\frac{H_{Z_{n-1}}}{P}$ in (\ref{2.22}), we have:
\begin{equation}
H_{XXXZ_{n-1}}-4H_{XZ_{n-1}}H_{X}-2H_{XX}H_{Z_{n-1}}=2(\beta_1)_X\quad .\label{2.30}
\end{equation}
We now define a new variable $Z_n$ such that:
\begin{equation}
\beta_1=H_{Z_n}\quad ,\label{2.31}
\end{equation}
which is again a hodograph transformation between the dependent variable $\beta_1$ and the independent one
$Z_n$. With this transformation, (\ref{2.30}) looks exactly like (\ref{2.29}) for $j=n$.
\begin{equation}
H_{XXXZ_{n-1}}-4H_{XZ_{n-1}}H_{X}-2H_{XX}H_{Z_{n-1}}=2H_{XZ_{n}}\label{2.32}
\end{equation}
Thus, by combining (\ref{2.29}) and (\ref{2.32}), we have the following $n-1$ CBS equations:
\begin{equation}
H_{XXXZ_j}-4H_{XZ_j}H_{X}-2H_{XX}H_{Z_j}=2H_{XZ_{j+1}},\quad j=1...n-1\quad .\label{2.33}
\end{equation}
\medskip

\noindent $\bullet$ Substitution of (\ref{2.31}) in (\ref{2.19}) gives us:
\begin{equation}
P_Y=PH_{XZ_n}-P_XH_{Z_n}\quad ,\label{2.34}
\end{equation}
whose compatibility with (2.28) yields:
\begin{equation}
H_{XXXZ_n}-4H_{XZ_n}H_{X}-2H_{XX}H_{Z_n}=2H_{XY}\quad ,\label{2.35}
\end{equation}
which is again a CBS equation in the variables $X$, $Z_n$ and $Y$.

\subsection{Summary of the transformations}
We  now summarize  the above results.

\medskip

Let us  start with the CHH(2+1)  system given by (\ref{2.11})-(\ref{2.14}). This is a system of $n$ fields
$V_1$,...$V_n$ and three independent variables:  $x$, $t$ and $y$. We have made the following
 transformations:

\medskip

1) \begin{equation}dX=Pdx+PV_ndt+\beta_1dy,\quad Z_1=t,\quad Y=y\quad .\label{2.36}\end{equation}

\medskip

2)
\begin{eqnarray}P_{XX}&=&2PH_X+\frac{P_X^2-1}{2P}\label{2.37}\\P_Y&=&PH_{XZ_n}-P_XH_{Z_n}\label{2.38}
\\P_{Z_1}&=&P^2(V_n)_X\quad .\label{2.39}\end{eqnarray}

\medskip

3) \begin{equation}H_{Z_{n-j}}=PV_j,\quad  j=1...n-1\quad .\label{2.40}\end{equation}
\medskip

4)  \begin{equation}H_{Z_n}=\beta_1 \quad .\label{2.41}\end{equation} With these transformations, we obtain
the following system:

\begin{eqnarray}H_{XXXZ_j}-4H_{XZ_j}H_{X}-2H_{XX}H_{Z_j}&=&2H_{XZ_{j+1}},\quad j=1...n-1\label{2.42}\\
H_{XXXZ_n}-4H_{XZ_n}H_{X}-2H_{XX}H_{Z_n}&=&2H_{XY}\quad .\label{2.43}\end{eqnarray}

We have now $n$ CBS equations for just one field $H$ and $n+2$ independent variables: $X$, $Y$,
$Z_1$...$Z_n$. It is fairly trivial to check that equations such as (\ref{2.42}) pass the Painlev\'e test
\cite{e04}. Consequently, the above described hodograph transformations, map the CHH(2+1) to a new system in
which the Painlev\'e techniques (Singular Manifold Method) can be applied. For $n=1$ it corresponds to the
result obtained by Hone in \cite{h00}.

\medskip

Notice that after the first reciprocal transformation, the system was (2.19)-(2.22) where obviously $P$,
$V_i$ and $\beta_1$ are considered as independent fields. The second and third hodograph transformations:
$$PV_j=H_{Z_{n-j}},\quad j=1...n-1$$
$$\beta_1=H_{Z_n}$$
imply that,  for any of the  $n$ independent fields $V_1,.....V_{n-1}$ and $\beta_1$ we define one of the $n$
variables $Z_1...Z_n$, that consequently are as independent as the $V_1,.....V_{n-1},\,\beta_1$ fields are.
Furthermore, in the appendix we will use the results of \cite{e04} to construct solitonic solutions  of
(\ref{2.42})-(\ref{2.43}) depending on the $n+2$ independent variables: $X$, $Y$, $Z_1$...$Z_n$. The main
benefit of the second and third hodograph transformations is that they allow us to write the equations in a
form in which the Lax pair can be algorithmically derived through the techniques of the singular manifold
method.

\medskip

We should remark that the hodograph transformation (\ref{2.36}) is not defined for peakons. Actually, as it
has been pointed in \cite{dhh02} and \cite{hw03}, (\ref{2.10}) breaks down when $U$ is a Dirac delta function
because the square root of a distribution is not defined.

\section{Integrability and Lax pair for CHH(2+1)}\label{lpCHH}
\setcounter{equation}{0} In a recent paper by us \cite{e04}, the singular manifold method \cite{Weiss} was
applied to CBS  to derive its Lax pair. By using these results,  the Lax pair for (\ref{2.42}) is:
\begin{equation} \psi_{XX}=\left(H_X+\frac{\lambda}{2}\right)\psi\quad .\label{3.1}\end{equation}
\begin{equation} 0=E_j=-\psi_{Z_{j+1}}+\lambda\psi_{Z_j}-H_{Z_j}\psi_X+\frac{H_{XZ_j}}{2}\psi,
\quad j=1...n-1\quad .\label{3.2}\end{equation} For  (\ref{2.43}), the spatial part is exactly the same, but
the temporal part is:
\begin{equation} 0=E_n=-\psi_{Y}+\lambda\psi_{Z_n}-H_{Z_n}\psi_X+\frac{H_{XZ_n}}{2}\psi\quad .\label{3.3}\end{equation}
Furthermore, the compatibility condition between (\ref{3.1}) and (\ref{3.2}) implies that the spectral
problem is non-isospectral because $\lambda$ satisfies:
\begin{equation} \lambda_X=0,\qquad \lambda_{Z_{j+1}}-\lambda\lambda_{Z_{j}}=0\quad .\label{3.4}\end{equation}
Analogously, the compatibility condition between (\ref{3.1}) and (\ref{3.3}) yields:
\begin{equation} \lambda_X=0,\qquad \lambda_{Y}-\lambda\lambda_{Z_{n}}=0\quad .\label{3.5}\end{equation}
Notice that (\ref{3.1}) is independent of the index $j$. Nevertheless, (\ref{3.2}) can be considered as a
recursion relation for the derivatives of $\psi$ with respect to each $Z_j$. This allows us to take the
following combination:
\begin{eqnarray} 0&=&E_n\lambda^{-n}+\sum_{j=1}^{n-1} E_j\lambda^{-j}\nonumber\\&=&-\lambda^{-n}\psi_{Y}+\lambda^{1-n}
\psi_{Z_n}-\lambda^{-n}H_{Z_n}\psi_X+\lambda^{-n}\frac{H_{XZ_n}}{2}\psi\nonumber\\&+&\sum_{j=1}^{n-1}
\left[-\lambda^{-j}\psi_{Z_{j+1}}+\lambda^{1-j}\psi_{Z_j}-\lambda^{-j}H_{Z_j}\psi_X+\lambda^{-j}\frac{H_{XZ_j}}
{2}\psi\right] \quad .\label{3.6}\end{eqnarray} It is easy to see that:
$$ \sum_{j=1}^{n-1}
\left[-\lambda^{-j}\psi_{Z_{j+1}}\right]+\sum_{j=1}^{n-1}
\left[\lambda^{-j+1}\psi_{Z_{j}}\right]=-\lambda^{1-n}\psi_{Z_n}+\psi_{Z_1}\quad .$$ Therefore, we have:
\begin{equation} 0=\psi_{Z_1}-\lambda^{-n}\psi_Y+\sum_{j=1}^n
\left[-\lambda^{-j}H_{Z_j}\psi_X+\lambda^{-j}\frac{H_{XZ_j}}{2}\psi\right]\quad .\label{3.7}\end{equation}
The combination of (\ref{3.4}) and (\ref{3.5}) gives us:
\begin{equation}  \lambda_{Y}-\lambda^n\lambda_{Z_{1}}=0\quad .\label{3.8}\end{equation}

\subsection{Inverse  transformation}
We can  now come back to the original fields $U$ and $V_j$ as well as to the original independent variables
$x$, $t$ and $y$. All  we need is to perform the change  \cite{h00}:
\begin{equation} \psi(X,Z_1...Z_n,Y)=\sqrt P\,\phi(x,t,y)\quad .\label{3.9}\end{equation}
And, according to  (\ref{2.17}), we have:

\begin{eqnarray}
\psi_X&=&\sqrt P\,\left(\frac{\phi_x}{P}+\frac{P_X}{2P}\phi\right)\nonumber\\\psi_{XX} &=&\sqrt
P\,\left(\frac{\phi_{xx}}{P^2}+\left[\frac{P_{XX}}{2P}-\frac{P_{X}^2}{4P^2}\right]\phi\right)\label{3.10}\\
\psi_{Z_1}&=&\sqrt P\,\left(\phi_t-V_n\phi_x+\frac{P_{Z_1}}{2P}\phi\right)\nonumber\\
\psi_{Y}&=&\sqrt P\,\left(\phi_y-\frac{\beta_1}{P}\phi_x+\frac{P_{Y}}{2P}\phi\right)\quad.\nonumber
\end{eqnarray} With these changes, (\ref{3.1}) becomes: $$
\frac{\phi_{xx}}{P^2}+\left[\frac{P_{XX}}{2P}-\frac{P_{X}^2}{4P^2}\right]\phi=
\left(H_X+\frac{\lambda}{2}\right)\phi\quad .$$ Or, by using (2.37) and (\ref{2.10}):
\begin{equation} \phi_{xx}=\left(\frac{1}{4}+\frac{\lambda}{2}U\right)\phi\quad ,\label{3.11}\end{equation}
which is the spatial part of  CHH(2+1). The temporal part can be obtained by using (\ref{3.10}) in
(\ref{3.7}). The result is: \begin{eqnarray}
0&=&\left[\phi_t-V_n\phi_x+\frac{P_{Z_1}}{2P}\phi\right]-\lambda^{-n}\left[\phi_y-\frac{\beta_1}{P}
\phi_x+\frac{P_{Y}}{2P}\phi\right]\nonumber\\&+&\sum_{j=1}^n
\lambda^{-j}\left[-H_{Z_j}\left(\frac{\phi_x}{P}+\frac{P_X}
{2P}\phi\right)+\frac{H_{XZ_j}}{2}\phi\right]\quad .\nonumber\end{eqnarray} We now need to use
(\ref{2.38})-(\ref{2.39})  to obtain:

\begin{eqnarray}
0&=&\left[\phi_t-V_n\phi_x+\frac{P(V_n)_X}{2}\phi\right]-\lambda^{-n}\left[\phi_y-\frac{\beta_1}{P}
\phi_x+\frac{PH_{XZ_n}-P_XH_{Z_n}}{2P}\phi\right]\nonumber\\&+&\sum_{j=1}^{n-1}
\lambda^{-j}\left[-H_{Z_j}\left(\frac{\phi_x}{P}+\frac{P_X}
{2P}\phi\right)+\frac{H_{XZ_j}}{2}\phi\right]-\lambda^nH_{Z_n}\left(\frac{\phi_x}{P}+\frac{P_X}
{2P}\phi\right)+\lambda^{-n}\frac{H_{XZ_n}}{2}\phi\quad ,\nonumber\end{eqnarray} which can be simplified
to:\begin{eqnarray} 0&=&
\phi_t-V_n\phi_x+\frac{P(V_n)_X}{2}\phi-\lambda^{-n}\left[\phi_y+\left(-\frac{\beta_1}{P}+
\frac{H_{Z_n}}{P}\right)\phi_x\right]\nonumber \\&+&\sum_{j=1}^{n-1}
\lambda^{-j}\left[-H_{Z_j}\left(\frac{\phi_x}{P}+\frac{P_X}
{2P}\phi\right)+\frac{H_{XZ_j}}{2}\phi\right]\quad .\nonumber\end{eqnarray} By using
(\ref{2.40})-(\ref{2.41}), we have: \begin{eqnarray}
0&=&\phi_t-V_n\phi_x+\frac{P(V_n)_X}{2}\phi-\lambda^{-n}\phi_y\nonumber
\\&+&\sum_{j=1}^{n-1} \lambda^{-j}\left[-V_{n-j}\left(\phi_x+\frac{P_X}
{2}\phi\right)+\frac{P(V_{n-j})_X+P_XV_{n-j}}{2}\phi\right]\quad ,\nonumber\end{eqnarray} and simplifying :

\begin{equation} \phi_t-\lambda^{-n}\phi_y-\left[V_n+\sum_{j=1}^{n-1}\lambda^{-j}V_{n-j}\right]
\phi_x+\frac{1}{2}\left[V_n+\sum_{j=1}^{n-1}\lambda^{-j}V_{n-j}\right]_x\phi=0\quad
.\label{3.12}\end{equation} Furthermore, by applying (2.18) to (3.8), we have the non-isospectral condition:

\begin{equation} \lambda_y-\lambda^{n}\lambda_t\quad =0.\label{3.13}\end{equation}

In sum: The Lax pair for the Hierarchy CHH(2+1) of equations in $2+1$ variables (\ref{2.7}) can be written
as:
\begin{equation} \phi_{xx}-\frac{\lambda}{2}U\phi=\frac{1}{4}\phi\label{3.14}\end{equation}
\begin{equation} \phi_t=\lambda^{-n}\phi_y +A\phi_x-\frac{A_x}{2}\phi\quad ,\label{3.15}\end{equation}
where
\begin{equation} A=\sum_{j=0}^{n-1}
\left[\lambda^{-j}V_{n-j}\right],\qquad \lambda_y-\lambda^{n}\lambda_t=0\quad .\label{3.16}
\end{equation}

\bigskip
 We have proved in \cite{e04} the usefulness of the Lax pair (\ref{3.1})-(\ref{3.2})
for solving CBS. Actually, in \cite{e04} we have used the singular manifold method to obtain Darboux
transformations for this Lax pair. These Darboux transformations are the basis for the construction of an
iterative and algorithmic procedure described in \cite{e04} that allows us to obtain a rich collection of non
trivial solutions. The inversion of the hodograph transformations (\ref{2.36})-(\ref{2.41}) provides us the
corresponding solutions for CHH(2+1) and its Lax pair. It will be the subject of future work.

\noindent{\bf\underbar {Remarks}}: Spectral problems similar to (3.10) have been considered in several papers
 \cite{ma80}, \cite{AF88}, \cite{cgp97}, \cite{hq02}, \cite{h00}. More precisely:

\medskip

$\bullet$ This Lax pair is  included in the scattering problem presented in equation (1.1) of reference
\cite{cgp97} and it corresponds to the case that these authors call the Dym case. Nevertheless, CHH(2+1) is
not included in  the cases that the authors presented explicitly because the generalization of the Camassa-
Holm hierarchy that they considered corresponds to $n=1$ (interchanging $t$ and $y$) and $U$ expanded as a
polynomial of degree $N-1$ in $\lambda$. Only the $n=1$ component of CHH(2+1) is equivalent to equation
(2.21) of \cite{cgp97} ($N=1$ case). The Lax pair for the $n=1$ component of the hyerarchie appears also in
\cite{h00}.

\medskip

$\bullet$ The Lax pair considered in \cite{hq02} for the negative Camassa-Holm hierarchy
(\ref{2.3})-(\ref{2.4}) can be obtained through the reduction $\frac{\partial}{\partial
y}=\frac{\partial}{\partial x}$. Equivalently, the Lax pair presented in the same reference \cite{hq02} for
the positive Camassa-Holm hierarchy arises from the reduction $\frac{\partial}{\partial t}=0$.  In our
notation, these $1+1$ Lax pairs are:

\begin{eqnarray} \phi_{xx}&=&\left(\frac{\lambda}{2}u+\frac{1}{4}\right)\phi\nonumber
\\ \phi_t&=& B\phi_x-\frac{B_x}{2}\phi\\B&=&\lambda^{-n}+\sum_{j=0}^{n-1}
\left[\lambda^{-j}V_{n-j}\right],\qquad \textrm{for}\, \textrm{the}\, \textrm{negative}\,
\textrm{hierarchy}\quad ,\nonumber
\end{eqnarray}
and:
\begin{eqnarray} \phi_{xx}&=&\left(\frac{\lambda}{2}u+\frac{1}{4}\right)\phi\nonumber
\\ \phi_y&=& C\phi_x-\frac{C_x}{2}\phi\\C&=&-\sum_{j=0}^{n-1}
\left[\lambda^{n-j}V_{n-j}\right],\qquad \textrm{for}\, \textrm{the} \, \textrm{positive}\,
\textrm{hierarchy}\quad .\nonumber
\end{eqnarray}

\medskip

$\bullet$ (3.18) corresponds  to  the $N=1$ (interchanging $t$ and $y$) case of \cite{AF88} (which
generalizes \cite{ma80}). (3.17) is not included in this reference because expansions in negative powers of
$\lambda$ were not considered there.
\section{Conclusions}
Here we have presented an extension of the $n$ component Camassa-Holm hierarchy to $2+1$ dimensions whose
$n=1$ component is a generalization of the Fokas-Fuchsteiner-Camassa-Holm equation. Although the Painlev\'e
test cannot be applied to this system, we have found a set of hodograph transformations that allows us to
transform the original CHH(2+1) into $n$ coupled CBS equations that pass the Painlev\'e test. This result
generalizes  \cite{h00} for a $n$ component hyerarchie. The relationship between integrable systems and the
Painlev\'e property is once again established.

CBS is known to have a non-isospectral Lax pair. This Lax pair was used in section 3 to invert the hodograph
transformations in order to obtain a non-isospectral lax pair for CHH(2+1). Note that the non-isospectral
condition $\lambda_y=\lambda^n\lambda_t$ depends on the order $n$ of the hierarchy.

The Lax pairs for the positive and negative  $1+1$ Camassa-Holm hierarchies can be obtained through the
reductions $\frac{\partial}{\partial t}=0$ and $\frac{\partial}{\partial y}=\frac{\partial}{\partial x}$
respectively.

\section*{Appendix}
There are a lot of solutions of the coupled CBS equations (2.42)-(2.43) that can be obtained by using the
techniques of \cite{e04}. The simplest solution can be constructed through the eigenfunctions of (3.1)-(3.3)
with $H=0$ and $\lambda$ constant. These eigenfunctions can be written as:
$$\psi=e^{(kX+\omega Y+\omega\sum_{j=1}^{n}\lambda^{(j-n-1)}Z_j)}\eqno(A.1)$$
where $\omega$ is a totally arbitrary constant and
$$k^2=\frac{\lambda}{2}$$
According to \cite{e04}, it allows us to construct the following singular manifold
$$\phi\sim 1+e^{2(kX+\omega Y+\omega\sum_{j=1}^{n}\lambda^{(j-n-1)}Z_j)}\eqno(A.2)$$
that yields to the following one-soliton solution:
$$H_{1-soliton}=-2\frac{\phi_{x}}{\phi}\eqno(A.3)$$
A two-soliton solution can be easily written by means of the same techniques of reference \cite{e04} (see
expressions (3.24)-(3.25) of this reference). The result is:
$$H_{2-soliton}=-2\frac{\tau_{x}}{\tau}\eqno(A.4)$$
where
$$\tau=\phi_1\phi_2-\Omega^2\eqno(A.5)$$
$\phi_1$ and $\phi_2$ are singular manifolds of the form (A.2) corresponding to two different spectral
parameters $\lambda_1$ and $\lambda_2$ and two different values $\omega_1$ and $\omega_2$ of $\omega$.
$$\Omega=\frac{\psi_1\psi_{2,x}-\psi_1\psi_{2,x}}{\lambda_2-\lambda_1}\eqno(A.6)$$
which implies:
$$\tau\sim 1+\psi_1^2+\psi_2^2+\left(\frac{k_1-k_2}{k_1+k_2}\right)^2\psi_1^2\psi_2^2\eqno(A.7)$$
where
$$\psi_1=e^{(k_1X+\omega_1 Y+\omega_1\sum_{j=1}^{n}\lambda_1^{(j-n-1)}Z_j)}\eqno(A.8)$$
$$\psi_2=e^{(k_2X+\omega_2 Y+\omega\sum_{j=1}^{n}\lambda_2^{(j-n-1)}Z_j)}\eqno(A.9)$$
and
$$k_1^2=\frac{\lambda_1}{2},\qquad k_2^2=\frac{\lambda_2}{2}$$

\section*{Acknowledgements}
This research has been supported in part by the DGICYT under projects BFM2002-02609
and \\ BFM2003-00078.

\end{document}